# THE BEHAVIOUR OF COPPER IN VIEW OF RADIATION DAMAGE IN THE LHC LUMINOSITY UPGRADE

R. Flukiger, T. Spina, CERN, Geneva, Switzerland

*Abstract*

In view of the safe operation of the quadrupoles in the luminosity upgrade of the LHC accelerator, the response of the copper stabilizer at low temperatures to the various high energy radiation sources is of primary importance. The present study takes into account the expected high energy spectrum of the simultaneous radiation by neutrons, protons, pions, electrons and photons, calculated using the FLUKA code by F. Cerutti (CERN) as well as on literature values. It was found that proton irradiation causes a considerably higher damage than neutron irradiation: in spite of a 3.8% proton fraction, the measured damage is of the order of 20%, which fits with the calculations of N. Mokhov (Fermilab) on the contribution of protons to the dpa. The same calculations indicate that the total effect of protons, pions and electrons is at least as high as that of neutrons. Since recent neutron experiments of Nakamoto et al. show that the RRR of Cu is reduced from 200 to 50-120 for a fluence of $10^{21}$ n/cm$^2$, it follows that the inclusion of all high energy sources would lead to RRR values well below 50, thus endangering stability and protection. This result confirms the necessity of including a tungsten shield inside the quadrupoles.

## INTRODUCTION

The operation of the luminosity upgrade of the LHC accelerator requires a rigorous control of the effects of high energy radiation on the performance of the quadrupole magnets. The stabilization problem is very complex, the quadrupoles being exposed to various radiation sources (photons, electrons, neutrons, protons and pions) over a broad energy spectrum reaching up to more than $10^3$ GeV [1]. A main difficulty resides in the fact that these high energy sources act *simultaneously* on the quadrupoles. Since there is at present no facility where the effect of more than one radiation source at a time can be studied, the only possible analysis consists in a separate study of the effects of each singular radiation source on the quadrupole components, i.e. the Nb$_3$Sn superconductor, the Cu stabilizer and the insulator. At present, it is not yet possible to predict whether the combined effect due to high energy radiation sources can be simulated by the simple sum of each one of them. Nevertheless, there are reasons to think that that this approximation may be correct, taking into account that the dpa (displacement per atom) induced by the combined radiations during the lifetime of the quadrupoles is quite low, of the order of $10^{-4}$.

The subject of the present paper is the description of the behavior of the Cu stabilizer present in the superconducting cable under the effect of the high energy irradiation in LHC Upgrade. Of particular interest is the enhancement of electrical resistivity, or in other words, the decrease of the Residual Resistivity Ratio (RRR). For Cu, the RRR is defined as the resistivity at room temperature divided by the one at 4.2 K, which is close to the 1.9 K operational temperature. The present study is based on a series of published works, relying on the radiation load inside of the quadrupole volume given by the calculations of F. Cerutti [1] with the FLUKA code [2]. These calculations show that there is a characteristic energy distribution for each one of the radiation sources mentioned above. The resulting radiation spectrum reaching the inner winding of Q2a (the quadrupole with the highest radiation load in the triplet, with an aperture of 150 mm) can be described by the fraction corresponding to each radiation source and the maximum energy of their respective distributions (Table 1).

Table 1: Radiation sources and track length fraction in LHC upgrade and approximate energies at the maximum of distribution

| Radiation source | Track length fraction (%) | Energy at distribution maximum (MeV) |
|---|---|---|
| Photons | 88 | ~ 0.5 |
| Electrons/positrons | 7 | ~1 – 10 |
| Neutrons | 4 | 1 |
| Pions | 0.45 | 100 - 200 |
| Protons | 0.15 | 100 - 200 |

The conditions inside the quadrupole winding vary for each one of the radiation sources. The gamma and neutron radiations will mainly traverse the magnet, in contrast to the charged particles: electrons and pions will be fully trapped. Protons show a more complex behavior: they will be almost fully trapped at energies below 100 MeV, where the penetration depth is of the order of the thickness of the inner winding (15 mm). The fact that the radiation sources acting on the quadrupoles have a wide distribution of energies renders the problem particularly difficult. Indeed, the presence of high energy gradients inside the windings leads to a highly inhomogeneous radiation damage, which will have strong consequences for the properties of the various quadrupole components, and thus also on the stabilizing Cu.

For LHC upgrade, the total absorbed radiation dose (in MGy = J/kg) has been calculated, as well as the number of dpa (displacements per atom), using the FLUKA code. The dpa value is not constant inside the inner coil of the quadrupole: a certain gradient is present as a consequence of the different particle energies and of the different penetration depths inside the coil. In order to guarantee a safe operation, one has thus to take into account the maximum load, and the values of importance are the peak dose where the effects are highest. From the calculations given in [1] it follows that the high energy fluences inside the second quadrupole (Q2a) winding for neutrons, protons and pions for the target integrated luminosity of 3000 fb$^{-1}$ will be approximately $2 \times 10^{21}$ neutrons/m$^2$ and $>10^{20}$ p/m$^2$ for both protons and pions.

## ELECTRICAL RESISTIVITY OF THE CU STABILIZER FOR DIFFERENT RADIATION SOURCES

A large quantity of data has been published in the last decades on the effects of the irradiation of Cu by various high energy sources and at various temperatures, from 4.2 K to room temperature. From the known literature data, the following conclusions can be drawn:

- The electrical resistivity $\rho(T)$ (and therefore the RRR value) of the Cu stabilizers in superconducting magnets is strongly affected by the various radiation sources (neutrons, protons and pions) The lowering of RRR after irradiation affects the quench stability and thus the protection scheme. The increase of resistivity in Cu can be explained by the generation of Frenkel pairs which change the scattering properties of conduction electrons: the electronic mean free path is reduced.
- The enhancement of electrical resistivity of Cu with irradiation depends strongly on the initial purity. The enhancement of resistivity (or decrease of the RRR) of Cu is considerably stronger for neutron irradiations at lower temperatures: the enhancement of electrical resistivity of high purity Cu (RRR $\geq$ 1'000) is of the order of 2 at room temperature and of ~ 5 at 77 K. Below 10 K, a factor of up to 50 was observed. In wires, with specified RRR values of the order of 150 - 200, the observed changes are considerably smaller.
- In order to be consistent with the behavior of magnets at 1.9 K, the following discussion will mainly consider data obtained on low temperature irradiations. After warming, a partial recovery of the RRR takes place: at 40 K, 30 - 50% of the RRR values are recovered (recovery stage I), depending on the initial state and purity of the Cu. Warming up to room temperature, the original RRR values are recovered up to ~ 90% (see Fig. 1).

Most known irradiation data on Cu been obtained on neutron irradiation. Due to the occurrence of various high energy radiation sources in LHC upgrade, it is important to evaluate the damage caused by each one of them. The energy loss due to atomic displacement as a particle traverses the material can be described by the Non-Ionizing Energy Loss (NIEL). The product of the NIEL and the particle fluence gives the displacement damage energy deposition per unit mass of material. This quantity has first been calculated for Si by van Ginneken [2] for various radiation sources, e.g. electrons, photons, neutrons, protons and pions. Si and GaAs are so far the only materials for which a detailed dependence of the NIEL for all these radiation sources has been published in detail. In the meantime, however, detailed calculations of the NIEL can already be performed by using the FLUKA transport code; this code has widely been applied for calculating the radiation load of the quadrupoles in LHC baseline and upgrade [1]. Although the conditions in Cu are very different from those encountered in Si, both material are crystalline, some general tendencies being expected to be similar. It appears from the calculations of van Ginneken [3] that for solid crystals, the NIEL by atomic displacement induced by electrons and photons is much smaller than that of neutrons, protons and pions, regardless of the particle energies. This tendency is also observed for the quadrupoles in the LHC upgrade.

### Irradiation by high energy electrons and photons

Electron bombardment introduces isolated simple defects. As the electron energy increases, the probability for two or more point defects to form a cluster increases. As mentioned above, the effects of electron and photon irradiation on Cu are expected to be considerably smaller than those of neutrons, protons and pions. A literature study has been performed, but no data were found for the effect of high energy photon irradiation of Cu, in contrast to the effect of electron irradiation on the electrical resistivity of Cu, which has been studied by several authors. Sassin [4] showed that for electron energies of 2.8 MeV the initial RRR ratio of 800 decreased by a factor of 38 for a fluence of $1.15 \times 10^{24}$ e/m$^2$. Taking into account that this fluence is > 100 times higher than the expected neutron fluence in LHC upgrade, the enhancement of the RRR ratio at equivalent electron fluences is expected to be considerably lower. The further considerations will be limited to the case of neutron and proton irradiation.

### Irradiation by high energy neutrons

As mentioned above, warming up to room temperature of Cu after neutron irradiations at T< 10K leads to strong recovery of the physical properties: several authors found a recovery of $\rho_o$ well above 90%. Guinan et al [5] and Horak et al. [6] have irradiated Cu samples characterized by RRR = 172 and 2'000, respectively. The irradiations were performed at 4.2 K with 14 MeV neutrons in the RTNS-II facility (fluence $10^{21}$ n/m$^2$) and at > 0.1 MeV (fluence $2 \times 10^{22}$ n/m$^2$), respectively. The results are shown in Fig. 1. It should be mentioned here that the superconductor Nb$_3$Sn shows a markedly different recovery behavior: the recovery at 300 K is only of the order of 10%, full recovery occurring above 500°C.

Stage I recovery is generally attributed to recombination of close Frenkel pairs and to the long-range migration of interstitials. In Cu the recovery mechanism is probably related to the local situation of the close Frenkel pairs, which may be separated by at least one stable lattice site beyond the spontaneous recombination radius (recombination will occur along the <110> or <100> directions). The total recovery in irradiated Cu is nearly identical for irradiations with 14 MeV neutrons (RTNS-II) and with reactor neutrons (> 0.1 MeV).

In a recent review article H. Weber [7] described the work of several authors on the behavior of the RRR of Cu after low temperature neutron irradiation. In particular, he mentioned the existence of a unique Kohler relation for Cu with RRR values close to the industrial ones. Kohler's rule states that the quantity $[\rho(B) - \rho(0)]/\rho(0)$ remains unchanged when increasing the impurity concentration $c$ and the field $B$ by the same factor. This dependence, i.e. $[\rho(B) - \rho(0)]/\rho(0)$ versus $B/\rho(0)$, where $\rho(0)$ is the zero-field resistivity, is important for predicting the evolution of $\rho$ under various field and irradiation conditions, even for different RRR values prior to irradiation.

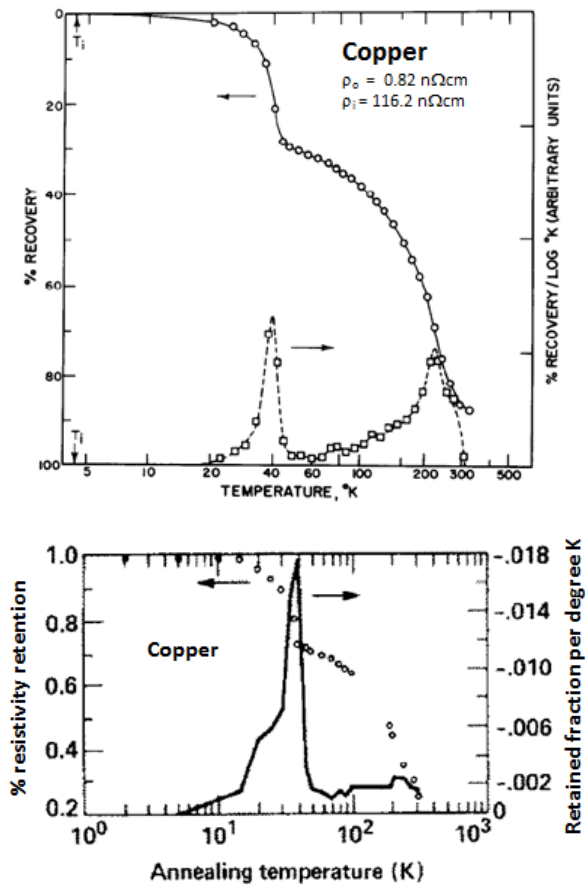

Figure 1: Post-irradiation, isochronal annealing results for Cu with different RRR values prior to irradiation, showing stage I recovery step at ~40 K. At 300 K, resistivity is recovered to 95% (Left: after Horak et al. [6], RRR = 2000; Right: after Guinan et al. [5], RRR = 100).

Recently, neutron irradiations on Cu have been performed at 14 K at KUR (Kyoto University Research Reactor Institute) at fluences up to of $2.8\times10^{21}$ n/m² by Nakamoto et al. [8]. As illustrated in Fig. 3, the electrical resistance increased proportionally with fluence, increasing from 2.1 to 3.05 µΩ (~50%). These data can be compared with the data of Horak et al. [5] shown in Fig. 1 by introducing the degradation rate $\Delta\rho/\phi t$, the resistivity enhancement for the applied fluence. In spite of the very different initial RRR ratios (RRR = 2000 for Horak et al. [6] and 300 for Nakamoto et al. [8]), the results are quite similar, the values for the degradation rate $\Delta\rho/\phi t$ being 0.58 and $0.82\times10^{-22}$ nΩm³, respectively. Nakamoto et al. [8] have measured the effect of neutron irradiation at 10 K on the RRR of superconducting wires after different fluences and found an essential effect: An initial RRR ratio of 200 would decrease to RRR = 160-190 for a neutron fluence of $10^{20}$ n/m² and to RRR = 50-120 for $10^{21}$ n/m², the latter being close to the total neutron fluence expected in LHC upgrade.

A last point concerning the variation of the RRR ratio of irradiated Cu concerns its behavior in presence of magnetic field. The only results treating the increase of stabilizer resistivity with high energy radiation neutron fluence have been discussed for neutron irradiation by H. Weber [7] and are reproduced in Fig. 4, where the ratio $\rho(B)/\rho_o(B)$ between the resistivity $\rho(B)$ after irradiation at 5 K and that one before irradiation $\rho_o(B)$ is plotted as a function of neutron fluence. It is seen that the enhancement of $\rho(B)/\rho_o(B)$ with fluence in the presence of magnetic field is considerably reduced, as a consequence of the decrease of magnetoresistivity for increasing zero-field resistance after irradiation.

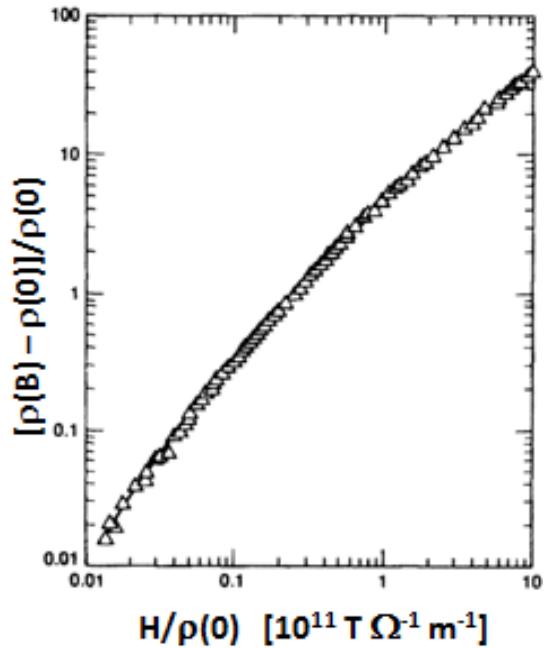

Figure 2: The Kohler plot for Cu samples for which the resistivity at zero field is entirely due to point defects (After Guinan et al. [5]).

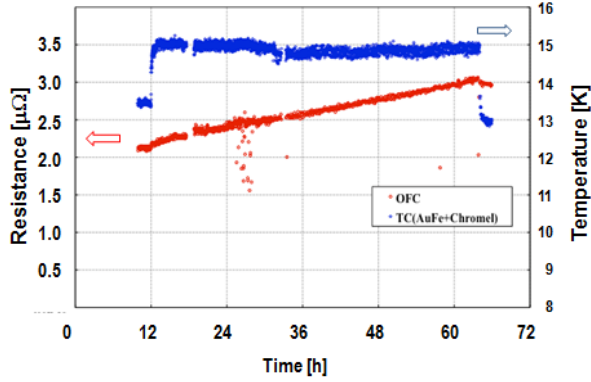

Figure 3: Enhancement of electrical resistance of Cu during irradiation at 14 K with fast neutrons (E>0.1 MeV). The temperature of 14 K was constant during the irradiation. This measurement has been extracted from the presentation of T. Nakamoto at the RESMM'12 [8].

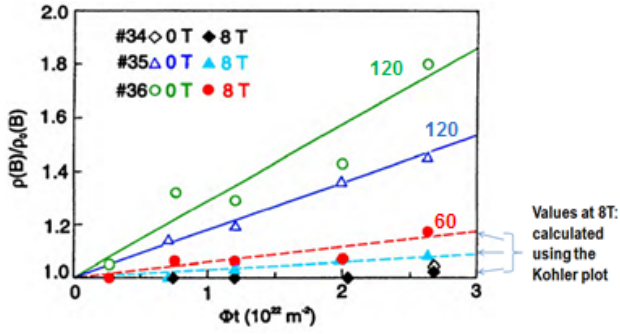

Figure 4: Increase of the electrical resistivity of the Cu stabilizer with neutron fluence (E > 0.1 MeV). The zero field data are measured, while the 8 T data were calculated from a Kohler plot. The data were obtained on Nb-Ti wires, but a similar effect is expected for $Nb_3Sn$ wires. The RRR values for the sample #34, #35 and #36 prior to irradiation are indicated in the figure (from H. Weber [7]).

*Irradiation by high energy protons*

In contrast to neutrons, protons and pions are charged particles, thus Coulomb elastic scattering will take place, which will considerably change the interactions with matter. Taking into account secondary ions, using the PHITS [9] and the MARS code, Mokhov [10] has calculated the contributions of the various high energy particles to the average dpa. He found that the major contributors to dpa (40%) are sub-threshold particles (particles with E < 100 keV + all fragments). Other important contributions were found to arise from neutrons (26%), protons (5%) and pions (15%) (see Table 2). A comparison between the contributions of protons and of neutrons to the total dpa values in Table 2 shows a ratio of ~ 20%, which is much higher than the corresponding fraction of ~ 4% resulting from Table 2: it follows that the relative effect of proton irradiation on the physical properties of the irradiated materials is expected to be larger than for neutron irradiation.

Table 2: Mean energy, flux and dpa, averaged over four hot spots (data from Mokhov [10])

| Particle j | <E> (GeV) | RMS (GeV) | Flux (cm$^{-2}$s$^{-1}$) | DPA/yr | DPA (%) |
|---|---|---|---|---|---|
| p | 2.93 | 10.7 | 1.3e8 | 1.75e-5 | 5 |
| n | 0.22 | 3.7 | 2.3e9 | 8.24e-5 | 26 |
| π, K | 13.8 | 41.6 | 5.4e8 | 4.78e-5 | 15 |
| μ | 11.3 | 19.7 | 6.3e5 | 1.70e-9 | - |
| γ | 0.018 | 0.35 | 8.6e10 | ~2.e-5 | 6 |
| e | 0.077 | 0.5 | 9.8e9 | 2.47e-5 | 8 |
| Sub-thresh. | | | | | 40 |

In order to describe the situation in the Cu stabilizer in the quadrupoles under the effect of protons with very different proton energies, it is important to present very recent dpa calculations on proton irradiated Cu, by Fukahori et al. [11]. These authors have used the PHITS code for taking into account the fact that high energy protons induce the formation of secondary ions, or in other words nuclear reactions occur before the stopping range (or penetration depth) of the protons in Cu is reached. The behavior at 14, 50, 200 and 200 MeV is shown in Fig. 5. These graphs illustrate quite well what is going on in the Cu stabilizer of the quadrupoles, where protons of all energies are present: the classical Bragg peak only occurs for 14 and 50 MeV protons, an increasingly different behavior being expected at higher energies. The secondary ions create new PKA's (primary knock-on atoms), which in turn result in enhanced dpa values. Taking into account that the thickness of the inner coil of the quadrupoles has a thickness of 15 mm, it follows that the protons at lower energies will have a strong contribution to the dpa values.

*Comparison between neutron and proton*

The question arises about the relative effects of neutrons and protons on the RRR of the Cu stabilizer in LHC upgrade. There is no direct comparison between the electrical resistivities of Cu stabilized $Nb_3Sn$ wires; the only comparative work was performed on the same Cu wire with a ratio RRR = 550 by Thompson et al. [12] (16 MeV protons at 4.2 K) and Roberto et al. [13] (15 MeV neutrons at 4.2 K). The RRR ratio is a factor ~ 2 higher than in $Nb_3Sn$ wires, but is not thought to influence the following conclusions. These authors [12] found that the damage caused by the different radiation sources (neutrons or protons) can be compared using their respective slopes $d\Delta\rho_o/d\Phi T$. From their results (Fig. 6), the ratio between the slopes for the neutron and the proton wires can be determined [12]:

$$\left.\frac{d\Delta\rho_0}{d\Phi T}\right|_n \left/ \left.\frac{d\Delta\rho_0}{d\Phi T}\right|_p \right. = 0.44$$

This ratio is approximately the same as the ratio between the corresponding damage energies cross sections $E_{DC}$, represented in Table 3.

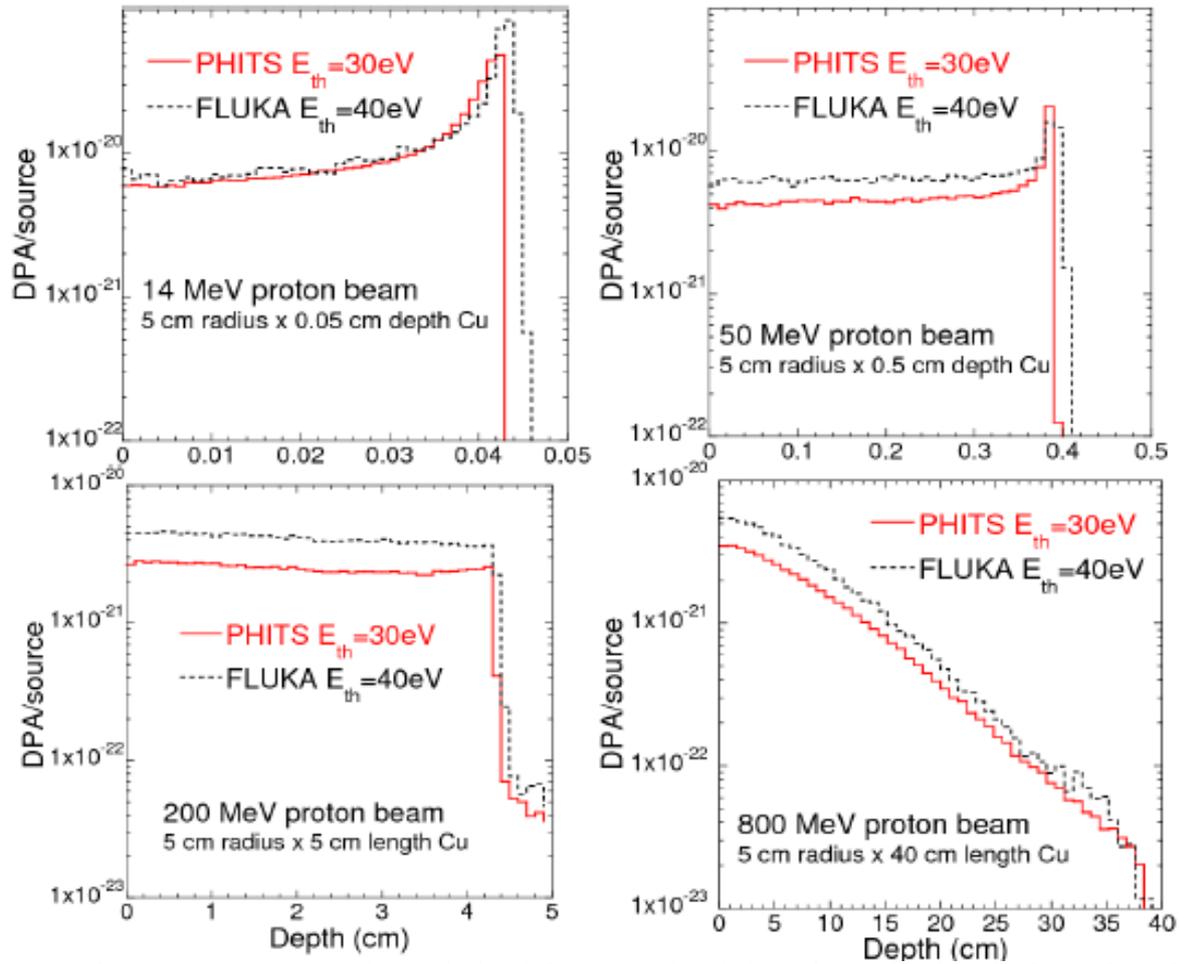

Figure 5: Displacements per atom (dpa) calculated for proton irradiation of Cu, showing the increasing effect of secondary ions with increasing energies. Both PHITS and FLUKA codes yield comparable results (From Fukahori et al. [10]).

Table 3: Damage energy cross sections for 16 MeV neutrons (d-Be) [13], for 15 MeV protons [12] and for E>0.1 MeV neutrons [7]

| Radiation source | Energy (MeV) | Damage energy cross sections $E_{DC}$ (keV.barn) | Ref. |
|---|---|---|---|
| neutrons (d-Be) | 15 | 263 | 13 |
| protons | 16 | 631 | 12 |
| neutrons | >0.1 | 78 | 7 |

It was recently shown [14] that the ratio between the initial slopes $dT_c/d\phi t$ of various Nb$_3$Sn wires irradiated in two different reactors, RTNS-II (14MeV) and TRIGA (>0.1MeV), correspond roughly to the ratio between the corresponding damage energy cross sections, $E_{DC}$(14MeV)/$E_D$(<0.1MeV). Since the neutron spectrum for neutrons in LHC Upgrade (neutron energies centered at 1 MeV) is similar to that in a TRIGA Mark-II reactor with > 0.1 MeV neutrons [14], the same argument can be applied here, and one can extend the above comparison to the estimated damage energy cross section of Cu in the TRIGA reactor.

Since only the damage energy cross sections for Ti and Nb have been calculated so far, the value of Cu has been taken as an intermediate value between both: $E_{DC}$(Nb$_3$Sn) ~ 75 keV · barn (see Table 3). The slope $d\Delta\rho_o/d\phi T$ corresponding to the lower damage energy for the 1 MeV neutrons has been plotted in Fig. 6. The present development constitutes only a first approximation, but indicates clearly that the same amount of damage on Cu wires (represented by $\Delta\rho/\rho$) produced by 16 MeV protons occurs at fluences a factor of 6 - 7 lower than for the 1 MeV neutrons produced in LHC Upgrade. It follows that in spite of the considerably lower proton fraction (3.8%) with respect to that of neutrons, the former induce a sizeable change of $\Delta\rho/\rho$ on the stabilizer, thus inducing an additional lowering of the RRR ratio. This confirms the dpa calculations of N. Mokhov shown in Table 2, where the percentage of total dpa caused by protons (Table 1) with respect to that of neutrons is 20%, i.e. much higher than the corresponding proton fraction.

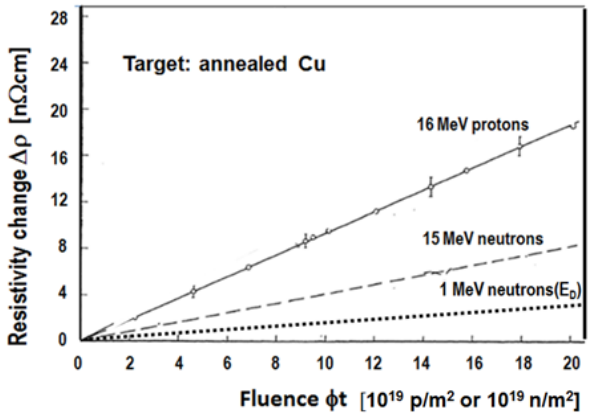

Figure 6: Enhancement of resistivity of Cu with RRR = 550 by 16 MeV neutrons [13] and by 15 MeV protons [12]. The dotted line at 1 MeV has been calculated assuming a ratio $E_{DC}(15MeV)/E_{DC}(1MeV) = 3.6$ [7].

*Irradiation by high energy pions*

Since there is no source with a sufficient pion flux to reach in reasonable time the fluences larger than the $10^{20}$ pions/m$^2$ expected in LHC upgrade, the effect can only be calculated. Supposing that the stopping range for the totality of pions falls inside of the quadrupole, Mokhov [10] calculated the contribution of pions on the total calculated dpa, based on the MARS code [15] and found ~ 15% (see Table 2). This is almost a factor three higher than the proton contribution. His results, listed in Table 2 for all high energy radiation sources, can now be used to get a rough estimation of the behavior of the Cu stabilizer in the quadrupoles. It follows that the total enhancement of the electrical resistivity of Cu due to the combined effect of protons and pions is almost as high as that one due to neutrons.

## CONCLUSIONS

Based on literature data, the effects of high energy irradiation on the Cu stabilizer in LHC upgrade have been briefly discussed. A common point to all radiation sources is the sizeable decrease of the RRR ratio with fluence, regardless of the high energy source and the irradiation temperature $T_i$. In contrast to the properties of Nb$_3$Sn, where there is little difference between irradiation at low temperature (T< 10 K) and 300 K, the properties of Cu (e.g. electrical resistivity, hardness, lattice parameter) exhibit an almost complete recovery when warming up to 300 K.

The present data allow an approximate view of the total damage of the various high energy sources in LHC upgrade on Cu, taking into account the effect of each single source on the dpa values as shown in Table 2. After a neutron fluence of $10^{21}$ n/m$^2$, Nakamoto [7] reported an already important effect, the starting ratio RRR = 200 being decreased to values between 50 and 120. Since the combined effect of protons and pions is almost as important as that one caused by neutron irradiation, this means that these values would be further lowered to extremely low values that could endanger stability and protection, even without taking into account the effect of the subthreshold particles (E < 100 keV). The effect of the latter is not known yet, but due to their high contribution to the dpa values (40%), an additional decrease has to be expected.

The question arises about the possibilities to maintain the RRR ratio at reasonable values. A periodic warming up the quadrupoles to 300 K could be considered, but the safer solution consists in inserting an internal shield to protect the quadrupole coils. This possibility is presently under study.